\documentclass[%
reprint,
superscriptaddress,
nofootinbib,
 amsmath,amssymb,
 aps,
 prl,
floatfix
]{revtex4-1}

\usepackage{graphicx}% Include figure files
\usepackage{dcolumn}% Align table columns on decimal point
\usepackage{bm}% bold math
\usepackage{paralist}
\usepackage[usenames,dvipsnames]{color}
\usepackage{blindtext}
\usepackage[separate-uncertainty = true,multi-part-units=single]{siunitx}
\usepackage{suffix}
\usepackage{siunitx}

\include{common_definitions}

\begin{document}

\title{Time-resolved inner-shell photoelectron spectroscopy:  from a bound molecule to an isolated atom}

\author{Felix Brau{\ss}e}
\author{Gildas Goldsztejn}
\affiliation{Max-Born-Institut f\"{u}r nichtlineare Optik und Kurzzeitspektroskopie, 12489 Berlin, Germany}
\author{Kasra Amini}
\affiliation{The Chemistry Research Laboratory, Department of Chemistry, University of Oxford, Oxford OX1 3TA, United Kingdom}
\author{Rebecca Boll}
\author{Sadia Bari}
\author{C\'{e}dric Bomme}
\affiliation{Deutsches Elektronen-Synchrotron DESY, 22607 Hamburg, Germany}
\author{Mark Brouard}
\author{Michael Burt}
\affiliation{The Chemistry Research Laboratory, Department of Chemistry, University of Oxford, Oxford OX1 3TA, United Kingdom}
\author{Barbara Cunha de Miranda}
\affiliation{Sorbonne Universit\'{e}s, UPMC Univ Paris 06, CNRS, UMR 7614, Laboratoire de Chimie Physique-Mati\`{e}re et Rayonnement, F-75005 Paris, France}
\author{Stefan D\"{u}sterer}
\author{Benjamin Erk}
\affiliation{Deutsches Elektronen-Synchrotron DESY, 22607 Hamburg, Germany}
\author{Marie G\'{e}l\'{e}oc}
\author{Romain Geneaux}
\affiliation{LIDYL, CEA, CNRS, Universit\'e Paris-Saclay, CEA Saclay, 91191 Gif-sur-Yvette, France}
\author{Alexander S. Gentleman}
\affiliation{The Physical and Theoretical Chemistry Laboratory, Department of Chemistry, University of Oxford, Oxford OX1 3QZ, United Kingdom}
\author{Renaud Guillemin}
\author{Iyas Ismail}
\affiliation{Sorbonne Universit\'{e}s, UPMC Univ Paris 06, CNRS, UMR 7614, Laboratoire de Chimie Physique-Mati\`{e}re et Rayonnement, F-75005 Paris, France}
\author{Per Johnsson}
\affiliation{Department of Physics, Lund University, 22100 Lund, Sweden}
\author{Lo\"{i}c Journel}
\affiliation{Sorbonne Universit\'{e}s, UPMC Univ Paris 06, CNRS, UMR 7614, Laboratoire de Chimie Physique-Mati\`{e}re et Rayonnement, F-75005 Paris, France}
\author{Thomas Kierspel}
\affiliation{Center for Free-Electron Laser Science (CFEL), Deutsches Elektronen-Synchrotron DESY, 22607 Hamburg, Germany}
\affiliation{Center for Ultrafast Imaging, Universit\"{a}t Hamburg, 22761 Hamburg, Germany}
\author{Hansjochen K\"{o}ckert}
\affiliation{The Chemistry Research Laboratory, Department of Chemistry, University of Oxford, Oxford OX1 3TA, United Kingdom}
\author{Jochen K\"{u}pper}
\affiliation{Center for Free-Electron Laser Science (CFEL), Deutsches Elektronen-Synchrotron DESY, 22607 Hamburg, Germany}
\affiliation{Center for Ultrafast Imaging, Universit\"{a}t Hamburg, 22761 Hamburg, Germany}
\affiliation{Department of Physics, Universit\"{a}t Hamburg, 22761 Hamburg, Germany}
\author{Pascal Lablanquie}
\affiliation{Sorbonne Universit\'{e}s, UPMC Univ Paris 06, CNRS, UMR 7614, Laboratoire de Chimie Physique-Mati\`{e}re et Rayonnement, F-75005 Paris, France}
\author{Jan Lahl}
\affiliation{Department of Physics, Lund University, 22100 Lund, Sweden}
\author{Jason W. L. Lee}
\author{Stuart R. Mackenzie}
\affiliation{The Physical and Theoretical Chemistry Laboratory, Department of Chemistry, University of Oxford, Oxford OX1 3QZ, United Kingdom}
\author{Sylvain Maclot}
\affiliation{Department of Physics, Lund University, 22100 Lund, Sweden}
\author{Bastian Manschwetus}
\affiliation{Deutsches Elektronen-Synchrotron DESY, 22607 Hamburg, Germany}
\author{Andrey S. Mereshchenko}
\affiliation{Saint-Petersburg State University, 7/9 Universitetskaya nab., St. Petersburg, 199034 Russia}
\author{Terence Mullins}
\affiliation{Center for Free-Electron Laser Science (CFEL), Deutsches Elektronen-Synchrotron DESY, 22607 Hamburg, Germany}
\author{Pavel K. Olshin}
\affiliation{Saint-Petersburg State University, 7/9 Universitetskaya nab., St. Petersburg, 199034 Russia}
\author{J\'{e}r\^{o}me Palaudoux}
\affiliation{Sorbonne Universit\'{e}s, UPMC Univ Paris 06, CNRS, UMR 7614, Laboratoire de Chimie Physique-Mati\`{e}re et Rayonnement, F-75005 Paris, France}
\author{Serguei Patchkovskii}
\affiliation{Max-Born-Institut f\"{u}r nichtlineare Optik und Kurzzeitspektroskopie, 12489 Berlin, Germany}
\author{Francis Penent}
\author{Maria Novella Piancastelli}
\affiliation{Sorbonne Universit\'{e}s, UPMC Univ Paris 06, CNRS, UMR 7614, Laboratoire de Chimie Physique-Mati\`{e}re et Rayonnement, F-75005 Paris, France}
\affiliation{Department of Physics and Astronomy, PO Box 516, Uppsala University, Uppsala, Sweden}
\author{Dimitrios Rompotis}
\affiliation{Deutsches Elektronen-Synchrotron DESY, 22607 Hamburg, Germany}
\author{Thierry Ruchon}
\affiliation{LIDYL, CEA, CNRS, Universit\'e Paris-Saclay, CEA Saclay, 91191 Gif-sur-Yvette, France}
\author{Artem Rudenko}
\affiliation{J.R. Macdonald Laboratory, Department of Physics, Kansas State University, Manhattan, KS 66506, USA}
\author{Evgeny Savelyev}
\author{Nora Schirmel}
\affiliation{Deutsches Elektronen-Synchrotron DESY, 22607 Hamburg, Germany}
\author{Simone Techert}
\affiliation{Deutsches Elektronen-Synchrotron DESY, 22607 Hamburg, Germany}
\affiliation{Max Planck Institute for Biophysical Chemistry, 33077 G\"{o}ttingen, Germany}
\affiliation{Institute for X-ray Physics, G\"{o}ttingen University, 33077 G\"{o}ttingen, Germany}
\author{Oksana Travnikova}
\affiliation{Sorbonne Universit\'{e}s, UPMC Univ Paris 06, CNRS, UMR 7614, Laboratoire de Chimie Physique-Mati\`{e}re et Rayonnement, F-75005 Paris, France}
\author{Sebastian Trippel}
\affiliation{Center for Free-Electron Laser Science (CFEL), Deutsches Elektronen-Synchrotron DESY, 22607 Hamburg, Germany}
\affiliation{Center for Ultrafast Imaging, Universit\"{a}t Hamburg, 22761 Hamburg, Germany}
\author{Jonathan G. Underwood}
\affiliation{Department of Physics and Astronomy, University College London, London WC1E 6BT, United Kingdom}
\author{Claire Vallance}
\affiliation{The Chemistry Research Laboratory, Department of Chemistry, University of Oxford, Oxford OX1 3TA, United Kingdom}
\author{Joss Wiese}
\affiliation{Center for Free-Electron Laser Science (CFEL), Deutsches Elektronen-Synchrotron DESY, 22607 Hamburg, Germany}
\author{Marc Simon}
\affiliation{Sorbonne Universit\'{e}s, UPMC Univ Paris 06, CNRS, UMR 7614, Laboratoire de Chimie Physique-Mati\`{e}re et Rayonnement, F-75005 Paris, France}
\author{David M. P. Holland}
\affiliation{Daresbury Laboratory, Daresbury, Warrington, Cheshire WA4 4AD, United Kingdom}
\author{Tatiana Marchenko}
\affiliation{Sorbonne Universit\'{e}s, UPMC Univ Paris 06, CNRS, UMR 7614, Laboratoire de Chimie Physique-Mati\`{e}re et Rayonnement, F-75005 Paris, France}
\author{Arnaud Rouz\'{e}e}
\email{rouzee@mbi-berlin.de}
\affiliation{Max-Born-Institut f\"{u}r nichtlineare Optik und Kurzzeitspektroskopie, 12489 Berlin, Germany}
\author{Daniel Rolles}
\affiliation{J.R. Macdonald Laboratory, Department of Physics, Kansas State University, Manhattan, KS 66506, USA}

\date{\today}% It is always \today, today,

\begin{abstract}

Due to its element- and site-specificity, inner-shell photoelectron spectroscopy is a widely used technique to probe the chemical structure of matter. Here we show that time-resolved inner-shell photoelectron spectroscopy can be employed to observe ultrafast chemical reactions and the electronic response to the nuclear motion with high sensitivity. The ultraviolet dissociation of iodomethane (CH$_3$I) is investigated by ionization above the iodine 4d edge, using time-resolved inner-shell photoelectron and photoion spectroscopy.  The dynamics observed in the photoelectron spectra appear earlier and are faster than those seen in the iodine fragments. The experimental results are interpreted using crystal field and spin-orbit configuration interaction calculations, and demonstrate that time-resolved inner-shell photoelectron spectroscopy is a powerful tool to directly track ultrafast structural and electronic transformations in gas-phase molecules. 
\end{abstract}

\maketitle

\section{\label{sec:intro}Introduction}

The observation of nuclear wave packet motion during molecular transformations represents a major step towards the understanding of molecular function and reactivity \cite{rlevine87:dynamics}, and is therefore actively pursued in experiments employing various time-resolved approaches. When a molecule, in its electronic ground state, is photoexcited through the promotion of an electron into an unoccupied orbital, complex reaction dynamics can take place, often involving the interplay between electronic and nuclear degrees of freedom, and the formation of intermediate products \cite{Worth,Yarkony,Levine}. Since the typical timescale for molecular vibrations is on the order of 10 to 100 fs, the direct observation of atomic motion during a photochemical reaction has only become possible with the development of femtosecond laser technologies. Pump-probe techniques \cite{fschem:Zewail}, using femtosecond lasers, have allowed ``images" of molecular structures at different stages of a reaction to be captured. More recently, the emergence of X-ray free-electron lasers (FELs) \cite{Emma,Ishikawa} and ultrafast relativistic electron pulse technologies \cite{Weathersby} have enabled time-resolved diffractive imaging studies on gas-phase molecules, and the first such experiments have demonstrated the possibility of visualizing directly the atomic motion with femtosecond temporal and {\AA}ngstrom-scale spatial resolution \cite{fel:Glownia,uelec:Yang2016}. 

As an alternative route, time-resolved photoelectron spectroscopy (TRPES) has been used extensively to investigate ultrafast molecular processes \cite{Neumark,stolow}. In these experiments, changes in the molecular structure are inferred from the angular and kinetic energy distributions of the photoelectrons emitted from the molecule by single or multiphoton ionization. While diffraction experiments are mainly sensitive to changes in the nuclear positions during a photochemical reaction, TRPES, which uses valence ionization by ultraviolet (UV) and extreme ultraviolet (EUV) laser pulses, is highly sensitive to the time evolution of the valence electronic structure, and can be used to investigate complex photochemical reaction processes involving intertwined electron-nuclear dynamics \cite{Blanchet,gessner,Bisgaard,Squibb2018}. Inner-shell ionization with X-rays offers similar insights into molecular structure and dynamics. Due to the strong localization of inner-shell orbitals, the transitions are element-specific and chemically selective, and inner-shell binding energies show characteristic chemical shifts that can provide a local probe of the environment of the ionized atoms \cite{Siegbahn}. Synchrotron radiation-based (soft) X-ray sources, in combination with photoelectron spectroscopy, have been widely used to investigate the static electronic and structural properties of isolated species, ranging from molecules to nanoparticles \cite{Morin,Miron}. Recently, femtosecond X-ray pulses have become available at large-scale facilities such as slicing synchrotron sources \cite{Schoenlein} and FELs \cite{Emma}, and several experiments have been proposed which aim to probe ultrafast molecular dynamics using time-resolved inner-shell electron spectroscopy. McFarland et al. \cite{McFarland} have reported time-resolved Auger electron spectroscopy experiments performed in UV photo-excited thymine molecules and a first attempt has been made recently at the LCLS free electron laser to observe changes to the carbon 1s photoelectron spectrum in UV-excited uracil \cite{Bolognesi}. However, to the best of our knowledge, no experiments have been reported that directly extract structural dynamics information using inner-shell photoelectrons as a probe.

Here, we present an experiment performed on CH$_3$I molecules undergoing ultrafast UV-induced dissociation probed by time-resolved soft X-ray inner-shell photoelectron and photoion spectroscopy. Our experimental results show that the time-resolved photoion spectra not only probe the dissociation dynamics but also contain information on additional processes, such as molecular Auger decays and charge transfer processes \cite{Erk}, induced by the probe pulse. These processes do not affect the fast photoelectrons, which can therefore be used to track directly the ultrafast structural transformations. Our experimental results are compared with theoretical predictions obtained from ab initio calculations modelling the excitation and subsequent decay process, and show a good agreement.
%figure1SIv2

\section{Methods}

\subsection{\label{sec:setup}Experimental setup}

\begin{figure}[h]
\centering
\includegraphics[scale=0.3]{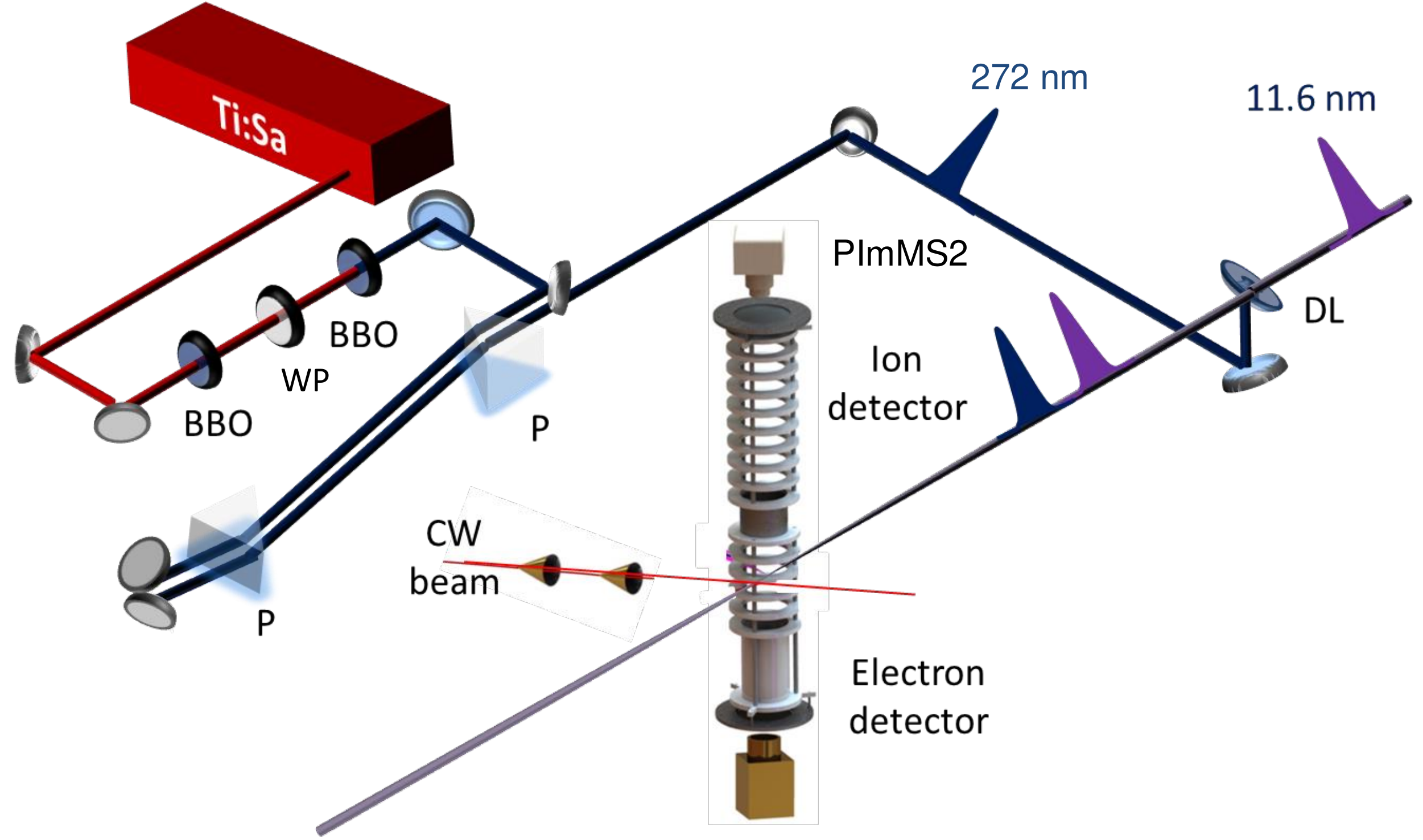} 
\caption{Sketch of the experimental setup. The 272 nm laser beam and the 11.6 nm FEL beam are collinearly overlapped and focused inside a target sample of iodomethane molecules at the center of a double-sided velocity map imaging spectrometer. Ion and electron momentum distributions are recorded at opposite ends of the spectrometer. The ion detector side is equipped with the PImMS2 camera that allows the arrival time and position of all ions to be recorded simultaneously while the electron detector incorporates a MCP/phosphor screen assembly followed by a CCD camera. P: prism; CW beam: Continuous molecular beam; DL: drilled mirror; BBO: Beta Barium Borate crystal; WP: wave plate.}
\label{fig:setup}
\end{figure}

The experiments were performed in the CAMP instrument installed at beamline BL 1 of the FLASH free-electron laser at DESY \cite{Feldhaus}. The experimental setup for such UV-pump, soft X-ray-probe experiments has been described previously \cite{Savelyev} and is only summarized here. During the beamtime, FLASH was operated in single-bunch mode to deliver ultrashort pulses of soft X-ray radiation at a central wavelength of 11.6 nm, with an average pulse energy of 115 $\mu$J at a 10 Hz repetition rate. The soft X-ray pulse duration was estimated around 120 fs full width at half maximum (FWHM). To reduce multiphoton interactions with the sample to a minimum, the FEL beam was typically attenuated with a 400 nm silicon filter, resulting in roughly 3 $\%$ transmission. The FEL pulse was collinearly overlapped with a 272 nm pump pulse obtained by third-harmonic generation of the 800 nm output pulse from the Ti:Sapphire pump-probe laser system at FLASH \cite{Redlin11} using a drilled mirror. A prism compressor installed in the 272 nm beam path was used to partially compress the UV pulse to 100 fs (FWHM). The maximum pulse energy of the UV pulse before the drilled mirror was 45 $\mu$J. The UV and the FEL pulses were focused inside a beam of CH$_3$I or CH$_2$ICl molecules, formed with a CW gas nozzle followed by two skimmers. The momentum distribution of the charged fragments (electrons and ions) resulting from the interaction of the molecules with the combined UV and FEL pulses was accelerated towards two position sensitive detectors facing each other using a double-sided velocity map imaging spectrometer (Fig. \ref{fig:setup}) \cite{Savelyev,Amini15}.  The electron momentum distributions were recorded using a 75 mm diameter chevron-pair MCP-detector followed by a phosphor screen (type P20) and a CCD,  while the ion momentum distributions were recorded using a 75 mm diameter chevron-pair MCP-detector followed by a fast phosphor screen (type P47) and the Pixel Imaging Mass Spectrometry (PImMS2) camera. This camera incorporates a detector array of 324x324 pixels for a time precision of 12.5 ns. Each pixel contains memory registers allowing the arrival time of up to four charged particles to be recorded per time-of-flight cycle \cite{Nomerotski10,John12,Brouard12,Amini15}. The camera was externally triggered to be synchronized to the 10 Hz repetition rate of the FEL. To correct for the inherent shot-to-shot fluctuations in the FEL parameters, single-shot electron and ion momentum distributions were recorded and post-processed later according to the procedure given in \cite{Savelyev}. 

\subsection{\label{sec:theory}Ab initio calculations}

All ab initio calculations were performed using the 3$^{rd}$ order Douglas-Kroll-Hess all-electron 2-component relativistic Hamiltonian \cite{Douglas74,Hess86,Fedorov03} as implemented in GAMESS-US \cite{Schmidt93,Gordon05}. The basis set of diffuse-augmented valence triple-zeta quality was used on iodine \cite{Patchkovskii06}, carbon \cite{deJong01}, and hydrogen \cite{deJong01}. With this basis set and minimal-CAS wavefunctions, the $^2$P$_{1/2}$ neutral state of iodine is calculated 0.845 eV above the ground $^2$P$_{3/2}$ state (compared to the experimental value of 0.946 eV \cite{Eppink99}). The first ionization potential of the iodine atom is underestimated by 0.23 eV, but the relative positions of the valence multiplet states agree with the experiment to better than 0.2 eV. For the 4d shell ionization, the atomic multiplet positions are systematically shifted by +4.35 eV, with the relative positions remaining in a good agreement with experiment, with errors not exceeding 0.3 eV (Table \ref{table1}).
\begin{table}
\caption{\label{table1} Lowest calculated per shell ionization potentials (IP) and relative atomic multiplet positions E$_{rel}$ within the shell for the low-lying states of I$^{+}$. Energies are in electron-volts. Experimental values are from \cite{OSullivan96}.}
 \begin{ruledtabular}
 \begin{tabular}{l|cccccc}
 J(5p) & J (4d) & J  & IP & E$_{rel}$ & IP & E$_{rel}$\\
 &  & (total) & calcd. &  calcd. & expt. & expt.\\ 
\hline
\hline
2 ($^{3}$P$_2$) & 0 & 2 & 10.21 & 0.00 & 10.45 & 0.00\\ 
0 ($^{3}$P$_0$) & 0 & 0 &       & 0.72 &       & 0.80\\ 
1 ($^{3}$P$_1$) & 0 & 1 &       & 0.68 &       & 0.88\\ 
2 ($^{1}$D$_2$) & 0 & 2 &       & 1.74 &       & 1.70\\ 
0 ($^{1}$S$_0$) & 0 & 0 &       & 3.76 &       & 3.66\\
\hline
3/2 ($^{2}$P$_{3/2}$) & 5/2 & 2 & 51.5 & 0.00 & 47.15 & 0.00\\ 
3/2 ($^{2}$P$_{3/2}$) & 5/2 & 3 &      & 0.08 &       & 0.17\\ 
3/2 ($^{2}$P$_{3/2}$) & 5/2 & 4 &      & 0.25 &       & -\\ 
3/2 ($^{2}$P$_{3/2}$) & 5/2 & 1 &      & 0.93 &       & 0.92\\ 
1/2 ($^{2}$P$_{1/2}$) & 5/2 & 2 &      & 1.20 &       & 1.36\\ 
1/2 ($^{2}$P$_{1/2}$) & 5/2 & 3 &      & 1.27 &       & 1.45\\ 
3/2 ($^{2}$P$_{3/2}$) & 3/2 & 2 &      & 1.69 &       & 1.80\\ 
3/2 ($^{2}$P$_{3/2}$) & 3/2 & 1 &      & 1.88 &       & 2.02\\ 
3/2 ($^{2}$P$_{3/2}$) & 3/2 & 3 &      & 2.22 &       & 2.21\\ 
3/2 ($^{2}$P$_{3/2}$) & 3/2 & 0 &      & 2.34 &       & 2.29\\ 
1/2 ($^{2}$P$_{1/2}$) & 3/2 & 2 &      & 2.56 &       & 2.88\\ 
1/2 ($^{2}$P$_{1/2}$) & 3/2 & 1 &      & 3.22 &       & 13.42\\ 
 \end{tabular}
 \end{ruledtabular}
 \end{table}

The geometry of the neutral ground state was optimized using scalar relativistic minimal-valence CASSCF(6,4) wavefunctions. Unconstrained geometry optimization leads to R(C-I)=2.169 {\AA}, R(C-H)=1.075 {\AA}, α(H-C-I)=107.3$^{\circ}$, compared to experimentally determined values of 2.134 {\AA}, 1.084 {\AA}, and 111.4$^{\circ}$, respectively \cite{Alekseyev07}. The dissociation of the C-I single bond was modelled by fixing the C-I distance between 2.0 and 5.0 {\AA} in 0.2 {\AA} increments, and optimizing the ground-state values for the remaining coordinates. At each geometry, the single-particle orbitals were optimized using a scalar-relativistic state-averaged minimal-valence CASSCF(6,4) wavefunction, using a dynamical-weighting window parameter of 5 eV \cite{Deskevich04}. Both singlets and triplets were included in the orbital optimization. 

\begin{figure}
\centering
\includegraphics[scale=0.5]{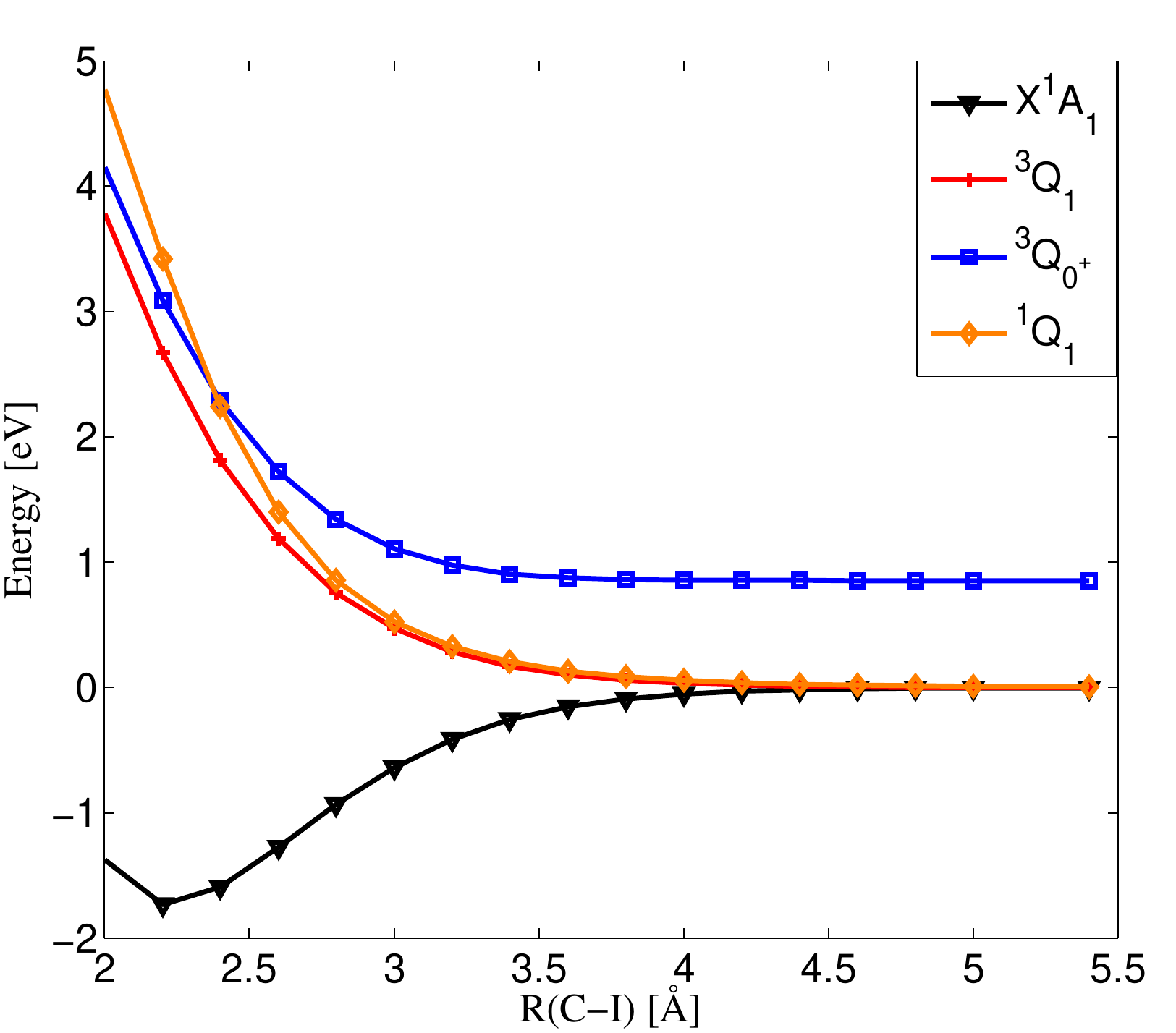}
\caption{Potential energy curves of the ground and selected neutral excited states of CH$_3$I as a function of C-I distance. The remaining structural parameters (R(C-H), α(H-C-I)) are optimized for the ground (X$^{1}$A$_1$) state. The symbols represent selected internuclear distances for which the calculation was performed.}
\label{fig:PES}
\end{figure}

Low-lying electronic states of the neutral molecules were calculated using spin-orbit configuration interaction (SO-CI) wavefunctions \cite{Fedorov03}, constructed from all minimal-valence CAS(6,4) determinants. The resulting low-lying electronic states (Fig. \ref{fig:PES}) are in a good agreement with the accurate ab initio results \cite{Alekseyev07}.

\begin{figure}
\centering
\includegraphics[scale=0.65]{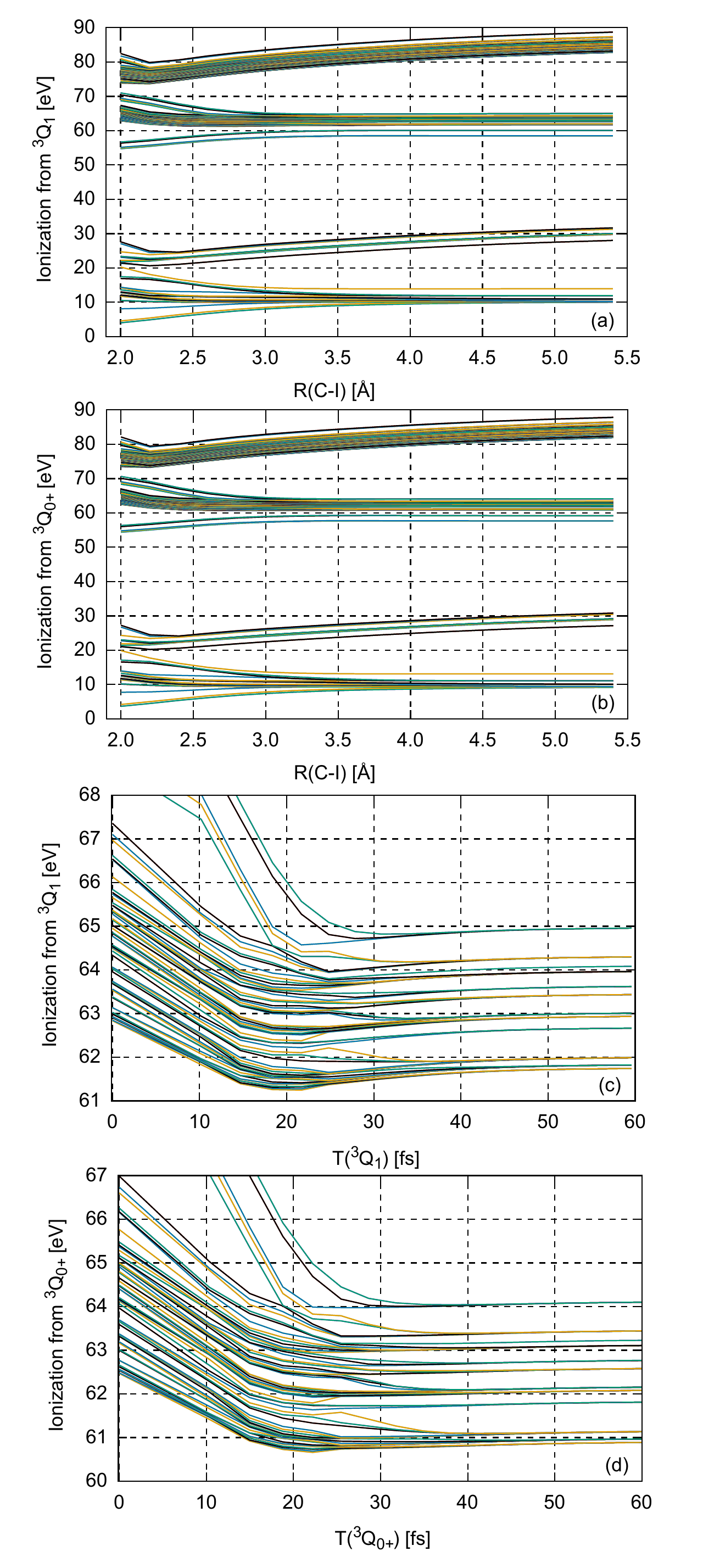}
\caption{Electronic structure of the CH$_3$I$^+$ molecular cation as a function of dissociation coordinate ((a)and (b)) or time after the initial excitation ((c)and (d)). Lines connect final cation states at each C-I distance ordered by energy, and do not imply continuity of electronic character of the state. Some of the final states may be inaccessible from the chosen initial state due to the selection rules, which were not taken into account. (a)and (b): calculated electron removal energy from the valence and I 4d shells as a function of distance, respectively. (c) and (d): 4d electron removal energies as a function of time after initial excitation. Panels: a) and c) the initial state is $^{3}$Q$_1$. b) and d) the initial state is $^{3}$Q$_{0^{+}}$.}
\label{fig:PESion}
\end{figure}

The relevant subset of the low-lying CH$_3$I$^{+}$ cation states was calculated from CAS(15,9) determinants, with the minimal-valence active space supplemented by the iodine 4d shell. Only the spin-free states with relative energies below 3.3 Hartree were included in the final SO-CI diagonalizations. This choice of the CI active space does not account for the electronic relaxation upon electron removal, leading to systematic shifts in the calculated multiplet energies involving each orbital shell (valence or 4d). 
Even for the small active space considered in our calculations, a very large number of final states arise due to the coupling between the two open shells in the cation. Four state manifolds are present in the calculation (Fig. \ref{fig:PESion}), namely: (1) single electron removal from the valence shell (the manifold converging to $\approx$ 10 eV); (2) valence electron removal accompanied by a valence excitation (the manifold converging to $\approx$ 30 eV); (3) single electron removal from the I 4d shell (the manifold converging to $\approx$ 60 eV); and (4) I 4d electron removal accompanied by a valence excitation (the manifold converging to $\approx$ 85 eV). Only the manifold converging to 60 eV is relevant to the interpretation of our experimental results. We emphasize that a large number of additional states will arise in this energy range in a calculation taking into account electron removal from the other occupied orbitals or excitations to low-lying Rydberg orbitals. However, such states are not relevant to the interpretation of our data.

Due to the large number of possible final states, the results of the SO-CI calculations are difficult to interpret. In order to develop a simpler, intuitively understandable model, we turn to the crystal-field theory \cite{Dunn65}, which has been used successfully to interpret the energy-level structure of iodine \cite{OSullivan96} and iodine-containing compounds \cite{Cutler92}. We adopt a model closely following the work of Cutler et al \cite{Cutler92}. Briefly, we consider the Hamiltonian $\hat{H}$ as a sum of an axial crystal-field Hamiltonian $\hat{H}_{CR}$ and a phenomenological spin-orbit Hamiltonian $\hat{H}_{SO}$:
\begin{equation} \label{eq1}
\begin{split}
& \hat{H}  = \hat{H}_{CR}+\hat{H}_{SO} \\
& \hat{H}_{CR} = 2\sqrt{\pi}V_0|\hat {Y}_{00}\rangle\langle \hat {Y}_{00}|+14\sqrt{\frac{\pi}{5}}V_2|\hat {Y}_{20}\rangle\langle \hat {Y}_{20}|\\
&+14\sqrt{\pi}V_4|\hat {Y}_{40}\rangle\langle \hat {Y}_{40}|\\
& \hat{H}_{SO} = \lambda_{SO}\hat{L}\cdot\hat{S}
 \end{split}
\end{equation}
where  $\hat{Y}_{LM}$ are spherical harmonics and $\hat{L}$  and $\hat{S}$ are respectively angular momentum and spin operators. The Hamiltonian acts within the Hilbert space consisting of the direct product of L=2 spatial and S=1/2 spin functions. The value of the spin-orbit coupling constant $\lambda_{SO}$ appropriate for CH$_3$I (0.695 eV) is taken from \cite{Cutler92}. The values of the crystal-field parameters $V_0$, $V_2$, and $V_4$ are determined by fitting the 4d orbital eigenvalues of the state-averaged scalar-relativistic Fock operator of the CASSCF calculation used to determine the SO-CI reference orbitals to the eigenvalues of $\hat{H}_{CR}$. This procedure uniquely defines the multiplet splitting parameters $V_2$ and $V_4$. The central position of the multiplet ($V_0$) is however determined with respect to the weighted average of the Fermi levels of the electronic states entering the Fock operator. Because the relative state energies and the state weights depend on the C-I internuclear separation, the coordinate-dependence of $V_0$ is not indicative of the absolute 4d removal energy from any specific state, and should be treated as somewhat arbitrary.  Finally, diagonalization of the full Hamiltonian yields the 1-particle energy levels, which can be used to estimate the dynamics of the I 4d lines during UV-dissociation. 
%At all distances, the multiplet structure is dominated by the spin-orbit coupling. As a result, the total angular momentum J labels can be assigned unambiguously, even though it is no longer a good quantum number of the axial CH$_3$I molecule.

\section{\label{sec:ion}Results and discussion}

\subsection{\label{sec:ion}Time-resolved ion measurements}
\begin{figure}
\centering
\includegraphics[scale=1]{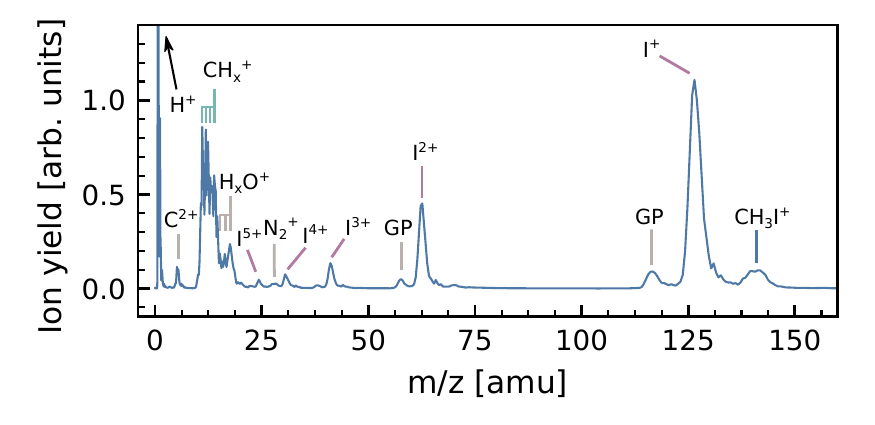}
\caption{Ion TOF spectrum of CH$_3$I following ionization by a 11.6 nm FEL pulse. I$^{n+}$ and CH$_x^{+}$ fragments are observed.  We note the presence of "ghost peaks", labeled as GP, that are observed when the ion drift tube voltage was higher than the front voltage on the MCP detector.}
\label{fig:TOF}
\end{figure}

The UV-induced dissociation of CH$_3$I has been studied extensively \cite{Eppink99,Nalda,Dura,Attar,Drescher}. The first absorption band (the $A$-band) arises from overlapping contributions of three dissociative electronic states (see Fig. \ref{fig:PES}), namely the $^{3}Q_{1}$ (E), $^{3}Q_{0^{+}}$(A$_1$), and $^{1}Q_{1}$ (E) states, that are dipole-allowed from the $^{1}A_{1}$ ground state. At 272 nm, the transition into the $^{3}Q_{0^{+}}$(A$_1$) state represents the major channel and leads to the formation of spin-orbit excited I$^{*}(^{2}P_{1/2})$ as the molecule dissociates. However, due to non-adiabatic couplings with the $^{1}Q_{1}$ (E) state along the C-I elongation coordinate \cite{Alekseyev}, population can be transferred to the $^{1}Q_{1}$ (E) state that converges towards the ground state I($^{2}P_{3/2})$ limit. 

While the UV pump pulse induces predominantly neutral dissociation, the soft X-ray probe pulse strongly ionizes the molecules via inner-shell ionization. A typical experimental ion time-of-flight (TOF) spectrum of CH$_3$I molecules exposed to the FEL pulse alone, recorded with the PImMS2 camera, is shown in Fig. \ref{fig:TOF}. The TOF spectrum contains I$^{n+}$ fragments and CH$_{x}^{+}$ fragments (where $x$ is the number of hydrogen atoms). Due to the giant $\epsilon$f$\leftarrow$ 4d centrifugal-barrier shape resonance \cite{Manson68} in iodine, the I 4d ionization cross-section at 107 eV (11.6 nm) is more than 10 times higher than that for valence ionization of CH$_3$I \cite{Olney98}, and is therefore the dominant ionization channel in our experiment. The vacancy in the inner-shell of the molecular ion relaxes within a few fs by one or two sequential Auger processes, leading to the formation of doubly and triply charged molecular ions. These ions finally fragment due to the fast charge redistribution of the positive charges that occurs throughout the molecular ion. The appearance of I$^{(3-5)+}$ ions in the TOF spectrum indicates that a second (or even a third) photon was absorbed by the molecular dications or trications within the 120 fs duration of the FEL pulse. Owing to the inherent increase in the internuclear distance that takes place following the absorption of the first photon, the charge redistribution becomes less efficient and most of the additional charges, due to the absorption of the second (and third) photon, remain on the multiply charged iodine ion \cite{Mertens16,Hollstein}.
\begin{figure}
\centering
\includegraphics[scale=1]{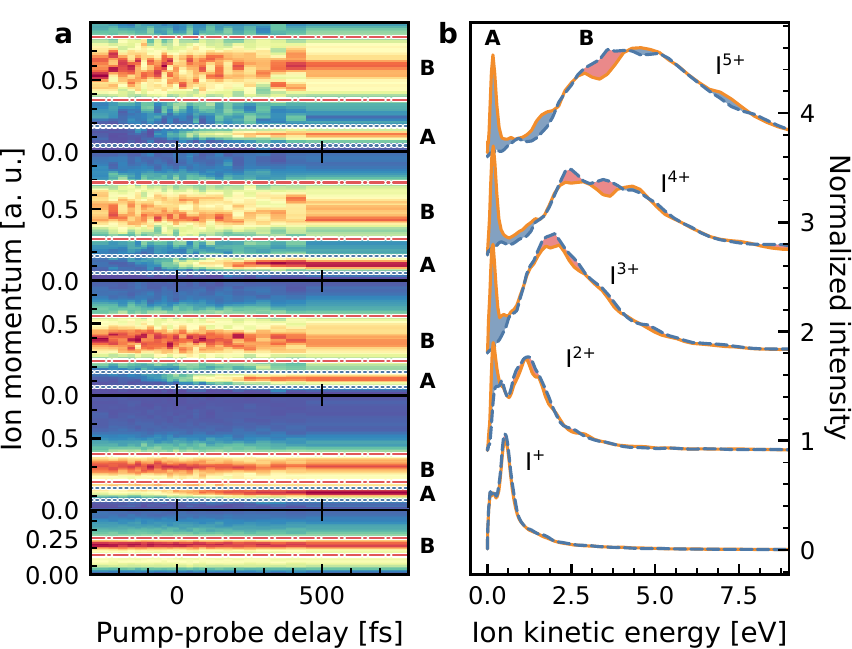}
\caption{Time-dependent I$^{n+}$ ion momentum distributions recorded in CH$_3$I (a) as a function of the UV pump-FEL probe delay and (b) the corresponding kinetic energy spectra for two delays: (dashed blue line) $\tau$= -1ps, i.e. FEL pulse comes first; (orange solid line) $\tau$ = +1ps, i.e. UV pulse comes first.  The blue and orange areas emphasize the increase and the depletion, respectively, of the signal when the UV pulse arrives before the FEL pulse. Channel A (dotted blue lines): ionization of the wave packet that propagates on the dissociative Q state manifold of CH$_3$I; Channel B (dash-dotted red lines): Coulomb explosion of the ground state molecules.}
\label{fig:TRion}
\end{figure}

As shown in previous studies \cite{Erk,Savelyev,Boll}, inner-shell photoionization of dissociating CH$_3$I molecules can result in low-energy, multiply charged iodine ions that appear in the time-resolved ion time-of-flight spectra when the UV pulse precedes the X-ray pulse. This effect is also observed in our experiment. Fig. \ref{fig:TRion} (a) displays ion momentum spectra, extracted from the PImMS2 camera for selected mass-over-charge ratios, as a function of the pump-probe delay between the 272 nm and the 11.6 nm pulses. The kinetic energy spectra, extracted before and after the time overlap, are also shown (Fig. \ref{fig:TRion}(b)). While the kinetic energy spectrum for singly ionized iodine is almost independent of the pump-probe delay, a sharp contribution appears at low kinetic energy in all I$^{n+}$ ion momentum distributions with n$>$1 when the UV pump pulse precedes the FEL pulse (labeled as A in Fig. \ref{fig:TRion}). The yield of this peak increases within the first few 100 fs following the time overlap, after which it remains constant. This contribution can be assigned to the ionization of the wave packet that propagates on the dissociative Q state manifold of CH$_3$I, leading to neutral CH$_3$ and I$^{n+}$ fragments. Therefore, this contribution reflects the translational kinetic energy that is acquired by the iodine fragment along the dissociative potential energy curves of the molecule following UV excitation. A second contribution at higher kinetic energies (labeled as B in Fig. \ref{fig:TRion}) is as well observed, which we assign to the Coulomb explosion of bound molecular ions (e.g. CH$_3$I$^{2+}$), following inner-shell ionization and molecular Auger decay. This contribution depends weakly of the time delay.

\begin{figure}
\centering
\includegraphics[scale=1]{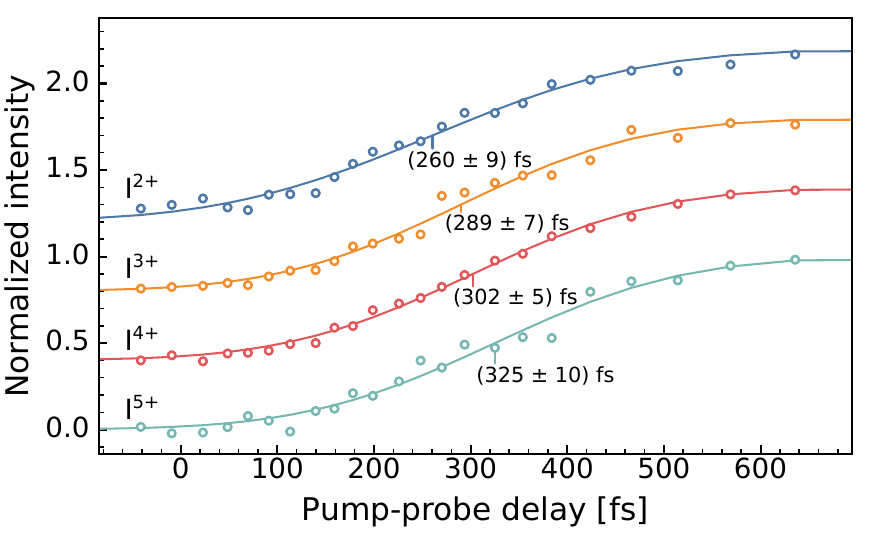}
\caption{Normalized integrated yield of the low-energy channel (integrated between 0 and 0.4 eV) in the multiply charged iodine ions plotted as a function of the delay between the UV pump and FEL probe pulses for several charge states of iodine (open circles), and the corresponding fit using a Gaussian cumulative distribution function (line). The centers of the fitted functions are indicated in parentheses, together with the standard deviations retrieved from the fits.}
\label{fig:TRion2}
\end{figure}

As previously stated, the delay-dependent channel that appears at low kinetic energy is due to ionization following UV-induced dissociation into neutral fragments. The maximum available energy, $E_{av}$(I), for the formation of I($^2$P$_{3/2}$) and I$^{*}$($^{2}$P$_{1/2}$) by UV-dissociation is given by:
\begin{equation}
E_{av}=\frac{m_{co-frag}}{m_{mol}}[h\nu-D_{0}-E_{so}-E_{i}^{mol}]
\end{equation}
with $h\nu$ being the excitation photon energy, $D_0$ the dissociation energy (2.41 eV for iodomethane \cite{Eppink99}), $E_{so}$(I) the spin-orbit splitting of atomic iodine (0.946 eV \cite{Eppink99}) and $E_i^{mol}$ the internal energy of the molecule. The quantity $m_{co-frag}$ is the mass of the co-fragment formed during neutral dissociation. Owing to the resolution of our velocity map imaging spectrometer for the voltage setting used here (50 meV for a kinetic energy below 1 eV), the two dissociative channels overlap in the final kinetic energy spectrum and cannot be distinguished. Previous measurements \cite{Murillo-Sanchez} have reported a quantum yield of 0.75 for the formation of I$^{*}$ in CH$_3$I at a photon energy of 266 nm. The kinetic energy of the delay-dependent channel is measured experimentally as 0.17 eV. This value is in close agreement with the expected maximum available energy, given by eq. (2), for the channels leading to the formation of I$^{*}$($^{2}$P$_{1/2}$) and I($^2$P$_{3/2}$) (0.13 eV and 0.23 eV, respectively). 

\begin{figure}
\centering
\includegraphics[scale=1]{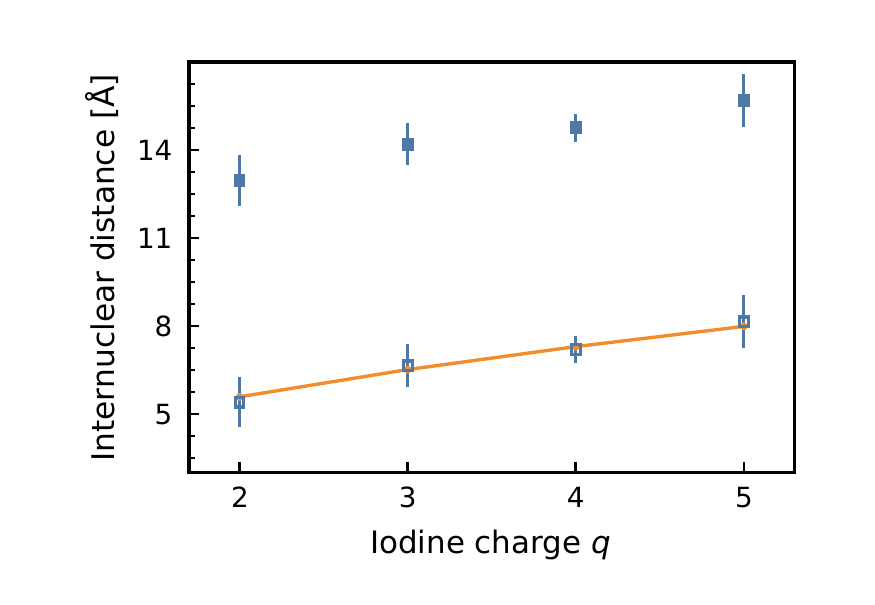}
\caption{Comparison between the critical internuclear distances predicted using eq. (3) (orange line) and the internuclear distances obtained from adiabatic propagation of the wavepacket on the $^{3}$Q$_{0^{+}}$ potential energy curve using the reaction times measured experimentally. The reaction times are used without including an additional time offset (full squares) and with an additional time offset (open squares).}
\label{fig:toffset}
\end{figure}

Interestingly, the onset of this low-energy channel has a specific, charge-state dependent delay due to intramolecular charge transfer that occurs following the removal of an initially localized inner-shell electron from the iodine atom in the course of the photodissociation \cite{Erk,Boll}. This is shown in Fig. \ref{fig:TRion2} together with the result of a fit using a Gaussian cumulative distribution function (CDF). The fitted parameters obtained from the measurement are summarized in Table \ref{table2}. Although the $\approx$ 220 fs width of the fitted CDF is independent of the charge state of the iodine ion, a clear shift of the center position occurs with increasing charge, in accord with the trend observed in previous experiments performed at higher photon energies \cite{Erk,Boll}. As the multiply ionized iodine atom separates from the methyl group, the Coulomb potential changes and the barrier between the two moving moieties increases. Therefore, the probability for electron transfer from the methyl group to the multiply ionized iodine atom decreases. At a certain critical internuclear distance, the barrier becomes higher than the binding energy of the highest occupied molecular orbital and charge redistribution is classically suppressed, finally leading to the observed channel.

The critical internuclear distance ($R_{cri}$) at which charge transfer is suppressed can be approximated well using a classical over-the-barrier model \cite{Boll}:
\begin{equation}
R_{cri}=\frac{(p+1)+2\sqrt{(p+1)q}}{E_i}
\end{equation}
with $p$ being the final charge state of the methyl group, $q$, the charge of the iodine atom and $E_i$ = 9.84 eV, the first ionization energy of the methyl group \cite{Schulenburg}. The critical internuclear distance obtained from eq. (3) can be compared to the internuclear distance that is expected from the dissociation of the molecule assuming adiabatic propagation of the wavepacket on the potential energy curve corresponding to the $^{3}$Q$_{0^{+}}$ state. 

Fig. \ref{fig:toffset} shows the expected internuclear distance (full squares in Fig. \ref{fig:toffset}) obtained from adiabatic propagation of the wavepacket at a time delay given by the center of the Gaussian cumulative function fitted to the experimental time-dependent iodine charge state yields (see Fig. \ref{fig:TRion2}), together with the result from the model given by eq. (3). A rather large discrepancy is observed. Since we were unable to observe a signal corresponding to the cross-correlation between the UV and FEL pulses in any of the measured fragments in our experiment, the absolute zero delay is not known precisely and therefore the fitted centers contain an additional delay t$_0$ that should be taken into account in the model. Using an additional delay t$_0$ as a fitting parameter, a rather good agreement can be obtained (open squares in Fig. \ref{fig:toffset}). The corresponding corrected reaction times extracted from this procedure are also given in Table \ref{table2}.

\begin{table}%[H] add [H] placement to break table across pages
\caption{\label{table2} Experimental centers and widths of the Gaussian cumulative distribution functions fitted to the delay-dependent ion yields shown in Fig. \ref{fig:TRion2}, along with the corresponding values obtained from the analysis of the delay-dependent photoelectron spectra.}
 \begin{ruledtabular}
 \begin{tabular}{l|ccc}
 Fragment & Center (fs) & Corrected center (fs) & Width (fs)\\
\hline
\hline
\\
I$^{2+}$ & 261$\pm$9 & 85 & 243$\pm$20 \\ 
I$^{3+}$ & 290$\pm$7 & 114 & 211$\pm$15 \\ 
I$^{4+}$ & 303$\pm$5 & 127 & 216$\pm$10 \\ 
I$^{5+}$ & 325$\pm$10 & 149 & 219$\pm$18 \\ 
e$^{-}$ molecule & 168$\pm$33 & -8 & 109$\pm$63 \\
e$^{-}$ atom & 195$\pm$42 & 19 & 125$\pm$80 \\
 \end{tabular}
 \end{ruledtabular}
 \end{table}

A quantitative analysis of the dissociation dynamics from this low kinetic energy channel in the fragment ion would require disentangling the influence of the distance-dependent charge transfer processes. As we show in the following, this feat can be avoided by analyzing the delay-dependence of the inner-shell photoelectrons emitted during the UV-induced dissociation.

\subsection{\label{sec:ion}Time-resolved photoelectron measurements}
\begin{figure}
\centering
\includegraphics[scale=0.90]{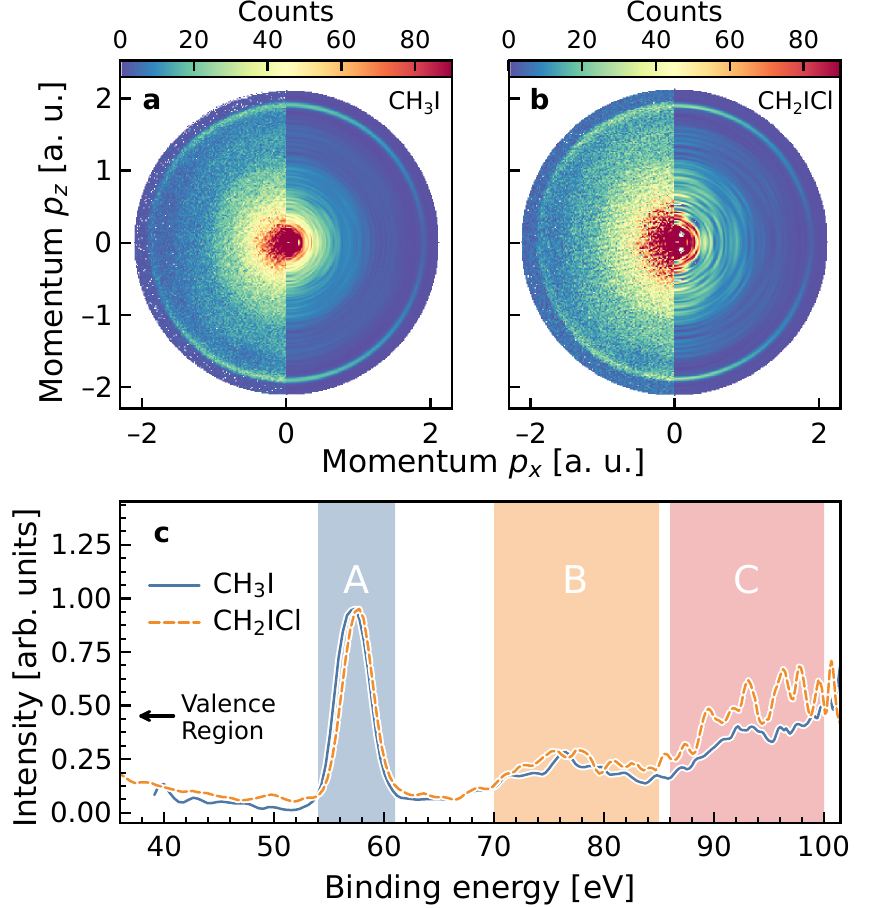}
\caption{Projected 2D electron momentum distributions (left half) and slices through the 3D photoelectron momentum distribution obtained after Abel inversion (right half) following ionization of CH$_3$I (a) and CH$_2$ICl (b) at a photon energy of 107 eV. (c) Corresponding electron spectra displayed as a function of the binding energy.}
\label{fig:electron}
\end{figure}

Slices through the three-dimensional photoelectron momentum distributions following inner-shell ionization of CH$_3$I, recorded simultaneously with the ion data discussed above, are displayed in Fig. \ref{fig:electron} together with the corresponding angle-integrated photoelectron kinetic energy spectrum (PES). At a photon energy of 107 eV, the spectra are dominated by the I 4d photoelectron peak near a binding energy of 57 eV (denoted as A in Fig. \ref{fig:electron} (c)). Additional contributions, denoted as B and C in Fig. \ref{fig:electron} (c), are assigned to Auger and shake-up electrons, as observed previously \cite{Holland}. We note that the calibration of the VMI detector was achieved in a separate measurement by recording the photoelectron momentum distribution of helium exposed to the 11.6 nm FEL pulse. 

\begin{figure}
\centering
\includegraphics[scale=1]{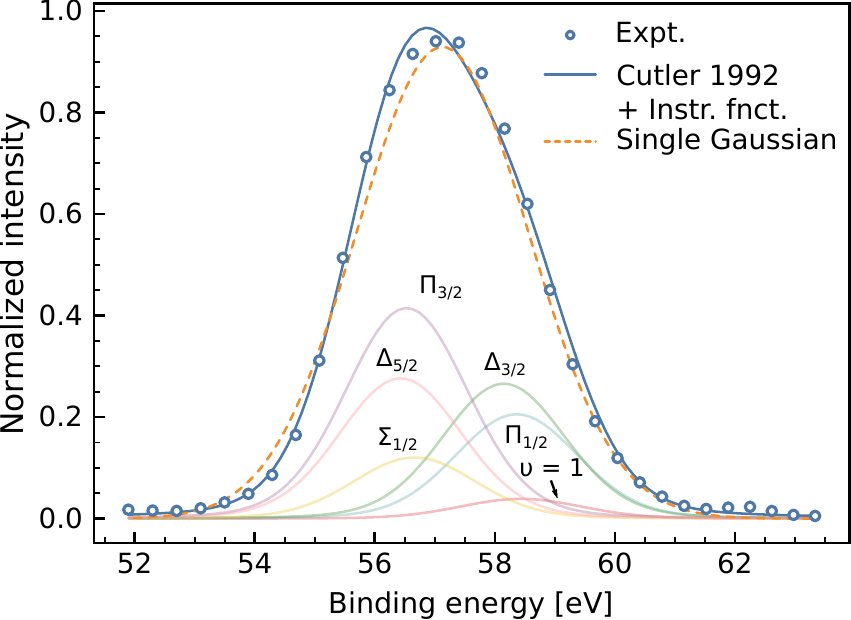}
\caption{Comparison between the experimental photoelectron spectrum measured in the CH$_3$I molecule following irradition by the 11.6 nm FEL pulse (open blue circle), and a convolution of the reported 4d photoline components ($\Sigma_{1/2}$, $\Delta_{5/2}$, $\Pi_{3/2}$, $\Sigma_{3/2}$, $\Pi_{1/2}$, $\nu=1$) with the experimental resolution (blue solid line). The experimental resolution is a sum of the instrumental function of the VMI spectrometer and the FEL bandwidth (~2.2 eV combined). The peak positions and Lorentzian widths are taken from Cutler et al. \cite{Cutler92} and are displayed in the figure. A comparison with a fit using a single Gaussian function is also included (dashed orange line).}
\label{fig:Cutler}
\end{figure}

An estimation of the energy resolution of the spectrometer was obtained by fitting the static photoelectron spectrum of CH$_3$I shown in Fig. \ref{fig:electron}. This spectrum was fitted by using Voigt profiles to represent the different ionization channels that contribute to the spin-orbit split iodine 4d photoelectron peaks. The Lorentzian widths and branching ratios of these contributions were kept fixed to the reported values \cite{Cutler92}. A Gaussian function was used to represent the instrumental resolution. In total, 6 channels were included in this fitting. These channels correspond to the contributions from the spin-orbit, ligand field, and vibrationally split states of the molecular cation that are formed following removal of an electron from the 4d shell (see Fig. \ref{fig:Cutler}). As a figure of merit we considered the $R^2$ measure of the fits. A $R^2$ of 0.999 was achieved for an instrumental Gaussian function with a 2.2 eV bandwidth (FWHM). The instrumental resolution is too low to resolve the 1.7 eV spin-orbit splitting of the I 4d photoline of CH$_3$I. We can, nevertheless, resolve a small absolute shift of 0.5 eV in the I 4d binding energy between CH$_3$I (peak position: 57.1 eV) and CH$_2$ICl (peak position: 57.6 eV), for which electron spectra were also recorded (see Fig. \ref{fig:electron} (c)). These values are in good agreement with the weighted average of the spin-orbit split 4d binding energies obtained from previous studies \cite{Novak,Nahon}, thereby demonstrating that our measurement is sensitive to shifts in the electron kinetic energy of a few hundred meV.

\begin{figure}
\centering
\includegraphics[scale=1]{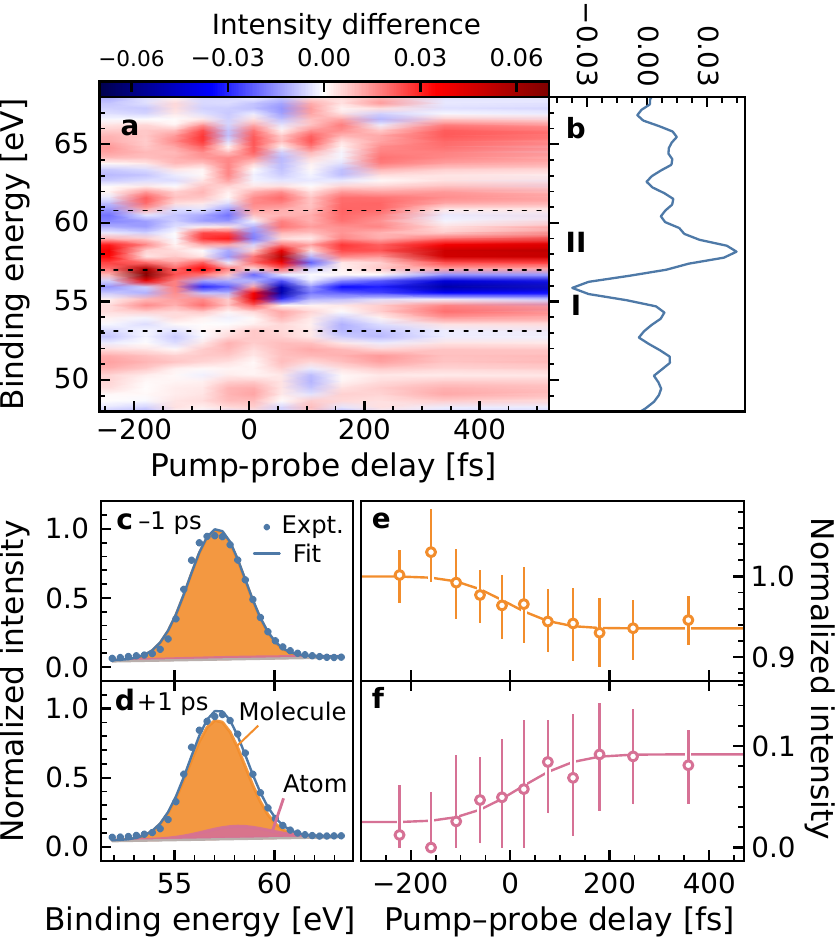}
\caption{(a) Time-dependent difference photoelectron spectra (see text) recorded in CH$_3$I as a function of the pump-probe delay. (b) Time-average difference spectrum calculated for positive delays near the iodine 4d line. Gaussian fits to the molecular and atomic components of the I 4d photoelectron peak, shown for time delays of -1 ps (c) and +1 ps (d). The symbols in panels (e) and (f) show the time-evolution of the intensity of the two Gaussians used to fit the depletion of the molecular 4d iodine contribution (e) and the rise of the atomic iodine 4d contribution (f). The lines are obtained from a fit using a cumulative Gaussian distribution function. The fit parameters are summarized in Table \ref{table2} along with the parameters obtained from the ion data.}
\label{fig:TRelectron}
\end{figure}

The change in the PES following UV excitation, as a function of the pump-probe time delay, is shown as difference spectra in Fig. \ref{fig:TRelectron}. At each time step, the PES recorded when the UV pulse is delayed by 1 ps with respect to the FEL pulse is subtracted from that recorded with the UV and FEL pulses, in order to emphasize the changes between excited and unexcited molecules. The most prominent delay-dependent effect is the appearance of a negative and a positive contribution to the I 4d photoelectron signal in the binding energy range 53-60 eV, labeled as regions I and II, respectively. The variation of the signal in this energy range is a consequence of the wave packet launched in the excited-state manifold of the CH$_3$I molecule by the UV pulse.  As the molecule evolves into a methyl radical and an isolated iodine atom, and the C-I distance increases in the dissociating molecule, the chemical shift of the 4d orbital decreases. This process is probed by the soft X-ray pulse, which ejects one electron from the I 4d shell into the continuum. The weighted average spin-orbit 4d binding energy in atomic iodine \cite{Nahon} is around 1.0 eV higher than that in CH$_3$I \cite{Novak}. Although the atomic and molecular components could not be completely separated in this experiment, we can, nevertheless, detect the resulting overall energy shift of the I 4d photoelectron line as a function of the delay. The drop of the signal in region I can therefore be associated with the depletion of intact molecules due to the dissociation, whereas the rise in region II can be attributed to the ionization of the iodine atoms that are formed. The oscillatory structure observed in the delay region between 0 and 500 fs is within the statistical uncertainty of the data and therefore cannot be interpreted further.
%The width of the Gaussian fitted to the molecular contribution is fixed to the width measured experimentally without pump pulse, whereas the width of the Gaussian representing the atomic contribution is used as a free parameter to account for a possible broadening of the atomic 4d photoline \cite{Cutler}.

To fully resolve the UV-dissociation dynamics in our time-resolved photoelectron measurements, a model that includes the spin-orbit splitting of the iodine 4d line for both the molecule and the atom would normally be required. The use of such a model would allow the energy of each component of the spin-orbit 4d molecular photoline to be fixed to the literature value. For free iodine atoms, the situation is more involved because the spectrum broadens due to open-shell couplings \cite{Nahon}. The modelling of this spectrum would require at least three components (Nahon et al. \cite{Nahon} use five transitions and Tremblay et al. \cite{Tremblay} present calculations with all 12 transitions).  Therefore, in total, we would need to fit the amplitude of at least five different contributions, the kinetic energies of the evolving spin-orbit split 4d atomic photoline, together with the respective widths. The number of fitting parameters is simply too large to be fitted reliably to our experimental data. Instead, we have used a simple model based on two Gaussian functions. The photoelectron spectrum recorded near the 4d iodine line can indeed be reasonably well approximated by a single Gaussian function, as shown in Fig. \ref{fig:Cutler}. This figure displays a comparison between the experimental photoeletron spectrum and the result of a fit using a single Gaussian function. In this case, a $R^2$ of 0.998 is achieved. At each time delay, the photoelectron spectrum was therefore fitted by the sum of two Gaussians. The first Gaussian was used to describe the contribution from the I 4d peak in CH$_3$I, whereas the second Gaussian was fitted to the contribution from the atomic iodine that was created after dissociation. The width of the Gaussian fitted to the molecular contribution was fixed to the width measured experimentally without pump pulse, whereas the width of the Gaussian representing the atomic contribution was used as a free parameter to account for a possible broadening of the atomic 4d photoline \cite{Nahon}. The peak positions of the two Gaussians were fixed to the weighted averages of the known values from measurements employing synchrotron radiation, i.e., to 57.3 eV for CH$_3$I \cite{Novak}, and to 58.3 eV for atomic iodine \cite{Nahon}. The amplitudes were used as fitting parameters. Note that this model does not take into account a dynamically shifting component as a function of pump-probe delay since our measurement has insufficient temporal and energy resolution to identify this component reliably. So, in total, three parameters were fitted.

%To quantify the dynamics observed in the PES, the signal near 55-60 eV was fitted by a sum of two Gaussians for each time delay, as illustrated in Figs. \ref{fig:TRelectron} (c) and (d). The first Gaussian describes the contribution from the I 4d peak in CH$_3$I, whereas the second Gaussian is fitted to the contribution from the atomic iodine that is created after dissociation (see SM for a detailed description).
% The peak positions of the two Gaussians are fixed to the weighted averages of the known values from measurements employing synchrotron radiation, i.e., to 57.3 eV for CH$_3$I \cite{Novak}, and to 58.3 eV for atomic iodine \cite{Nahon}. The amplitudes are used as fitting parameters. Note that this model does not take into account a dynamically shifting component as a function of pump-probe delay since our measurement has insufficient temporal and energy resolution to identify this component reliably.
The results of the fitting procedure are presented in Figs. \ref{fig:TRelectron} (e) and (f) by plotting the intensity of the two Gaussian functions described above as a function of pump-probe delay. These time-dependent intensities are subsequently fitted with a Gaussian cumulative distribution function. Comparing the widths and positions of the CDFs fitted to the electron data with those obtained from the ion data, all of which are summarized in Table \ref{table2}, a significant difference in the response of the electrons and the ions to the UV-induced dissociation is evident. The depletion of the I 4d contribution in CH$_3$I, centered at (-8$\pm$33) fs, and the appearance of the 4d atomic photoline ($19\pm 42$) fs coincide in time with the arrival of the UV pulse and occur with a decay time (molecular contribution) and rise time (atomic contribution) of $\approx$ 120 fs. This is a much faster and narrower onset than that of the low-energy channel in the fragment ions. This indicates that the inner-shell photoelectrons are a much more direct probe of the changes in the molecular electronic and nuclear structure occurring during the dissociation than are the fragment ions, the latter being affected by Auger decay and charge redistribution processes that occur over a more extended period of time. Remarkably, we find that the electronic structure in the free atom, as measured by the inner-shell photoelectrons, is established faster than the time resolution of our experiment, consistent with findings from transient absorption spectroscopy \cite{Drescher}.
\begin{figure}
\centering
\includegraphics[scale=0.55]{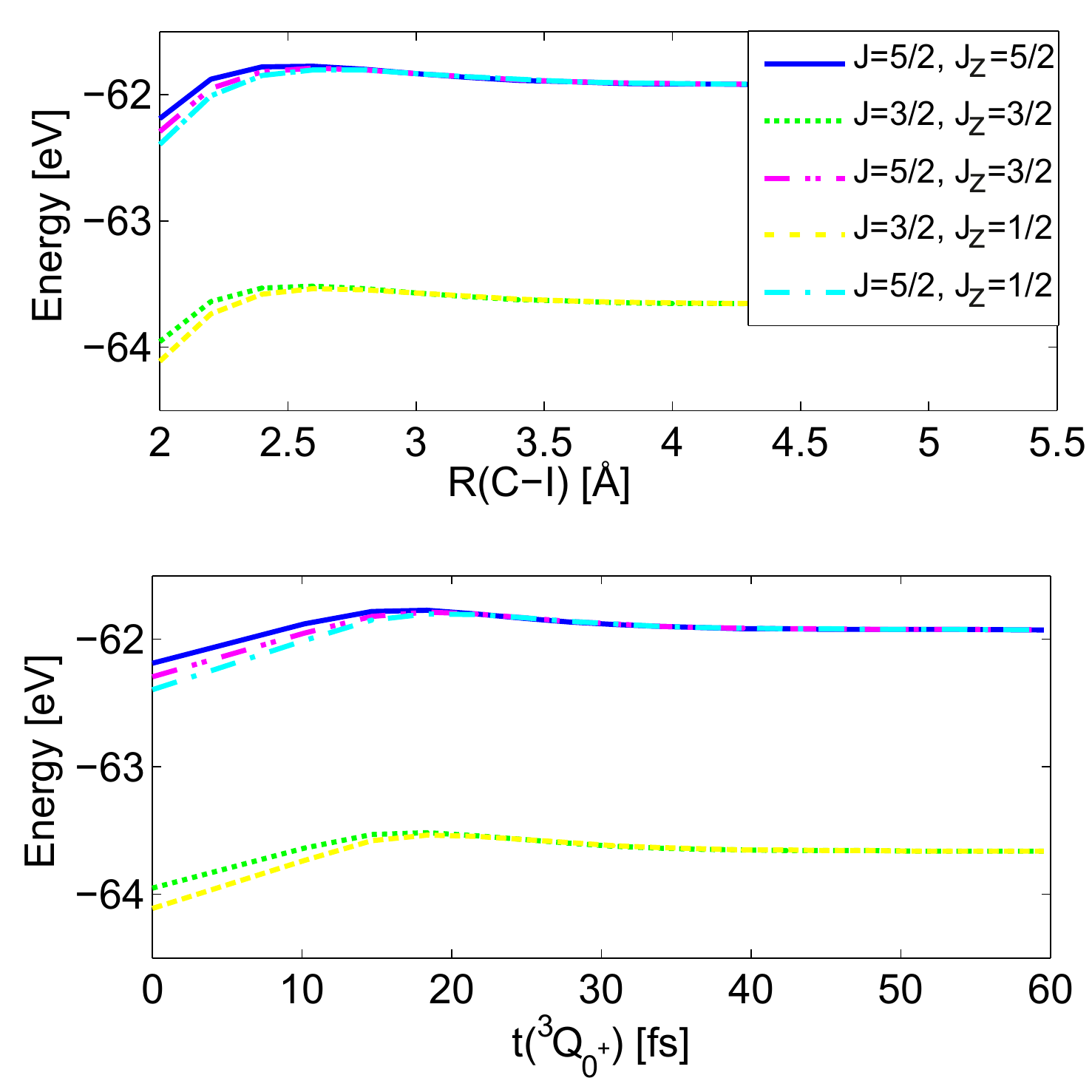}
\caption{(Upper panel) Calculated evolution of the crystal-field binding energy of the I 4d$_{5/2}$ and 4d$_{3/2}$ states along the C-I bond coordinate.  J and J$_{z}$ are the total angular momentum and the component of the total angular momentum along the principle molecular axis, respectively. (Lower panel) Time evolution of the binding energy of the I 4d$_{5/2}$ and 4d$_{3/2}$ states, respectively, along the $^{3}Q_{0^{+}}$ dissociation pathway, treated classically. Similar results are observed for the $^{3}Q_{1}$ state. Absolute binding energies are overestimated due to the incomplete treatment of the electron-shell relaxation.}
\label{fig:calculation}
\end{figure}

To further elucidate the time scale of the expected change in the chemical shift of the I 4d levels and to corroborate our interpretation, we have performed two different calculations. The dependence of the energy of the core-excited states on the C-I coordinate is first estimated using a model based on the crystal field theory  \cite{Cutler92} (see Methods section). The result of this procedure is shown in Fig. \ref{fig:calculation}. We observe a rapid change in the binding energy of the iodine 4d$_{5/2}$ and 4d$_{3/2}$ manifold, which occurs within the first $\approx$ 20 fs following the UV excitation. The atomic limit is reached after $\approx$ 40 fs. The crystal field model underestimates the total change in the binding potential due to the arbitrariness in the choice of the Fermi level in this calculation. More accurate calculations based on the SO-CI method using an active space including the 4d orbitals were therefore performed to estimate the potential energy curves of the CH$_3$I molecular cation near the I 4d ionization energy (see Fig. \ref{fig:PESion}). A qualitatively similar behavior is observed. For the $^{3}Q_{0^{+}}$ and $^{3}Q_{1}$ dissociative pathways, we calculate a binding energy shift of 1.3 eV, which is close to the experimental value. This energy shift appears within the first $\approx$ 20 fs of the dissociation, reaching an asymptotic value near $\approx$ 45 fs, which is consistent with our experimental observation.
%An energy shift of the 4d photoelectron line is observed as a function of pump-probe delay and attributed to a changing chemical shift in the dissociating molecule,
%We observe that the dynamics extracted from the electron spectra appear earlier and are faster than those associated with the I$^{n+}$ ions. Since the fast photoelectrons emitted during dissociation are not affected by charge transfer processes occurring in the molecular ions that are left behind, they can be used to directly track ultrafast structural transformations and to quantify ultrafast molecular rearrangements.  
%\section{Conclusions}

\section{\label{sec:ion}Conclusion}

In conclusion, we have investigated the UV-induced dissociation dynamics of CH$_3$I using femtosecond time-resolved inner-shell photoionization spectroscopy above the I 4d edge. A reduction in the intensity of the I 4d peak from CH$_3$I is observed together with the appearance of a new contribution attributed to ionization of the iodine atoms that are formed by photodissociation. This experimental evidence can be used to trace the transition from a bound molecule to an isolated atom.  While the temporal resolution of the current experiment was insufficient to fully resolve this process, which is predicted to occur within $\approx$ 40 fs, the development of sources delivering ultrashort pulses of short-wavelength radiation, e.g., based on high-order harmonic generation together with a time-delay compensating monochromator, allowing tunable sub-20 fs, narrowband ($<$ 500 meV) soft X-ray pulses \cite{Eckstein} to be generated, opens up this prospect. Also, inner-shell TRPES can benefit significantly from the use of a seeded FEL, such as FERMI, or self-seeding technologies that allow the temporal coherence to be improved and to obtain almost Fourier-transform limited XUV and X-ray pulses. Such sources, combined with high-resolution photoelectron spectroscopy, can become a powerful tool for exploring ultrafast molecular dynamics. 

\section{\label{sec:ion}Aknowledgement}

\begin{acknowledgments}
We gratefully acknowledge the work of the scientific and technical team at FLASH, who has made these experiments possible. The support of the UK EPSRC (to M.Br., S.R.M. and C.V. via Programme Grants EP/G00224X/1 and EP/L005913/1), the EU (to M.Br. via FP7 EU People ITN project 238671 and to J.K., P.J., J.L., H.S, and D.R. via the MEDEA project within the Horizon 2020 research and innovation programme under the Marie Skłodowska-Curie grant agreement No 641789), STFC through PNPAS award and a mini-IPS grant (ST/J002895/1), and a proof of concept grant from ISIS Innovation Ltd. are gratefully acknowledged. We also acknowledge the Max Planck Society for funding the development of the CAMP instrument within the ASG at CFEL. In addition, the installation of CAMP at FLASH was partially funded by BMBF grant 05K10KT2. K.A. thanks the EPSRC, Merton College, Oxford University and RSC for support. A.Ru. and D.R. acknowledge support from the Chemical Sciences, Geosciences, and Biosciences Division, Office of Basic Energy Sciences, Office of Science, U.S. Department of Energy, Grant No. DE-FG02-86ER13491. D.R., E.S., R.B., C.B, S.B. and B.E. were also supported by the Helmholtz Gemeinschaft through the Helmholtz Young Investigator Program. J.K. was, in addition to DESY, supported by Helmholtz Networking and Initiative Funds, by the excellence cluster ``The Hamburg Center for Ultrafast Imaging-Structure, Dynamics and Control of Matter at the Atomic Scale'' of the Deutsche Forschungsgemeinschaft (CUI, DFG-EXC1074), and, with T.K., by the Helmholtz Virtual Institute 419 “Dynamic Pathways in Multidimensional Landscapes”. T.R. and R.G. acknowledge the French Agence Nationale de la Recherche (ANR) through XSTASE project (ANR-14-CE32-0010). T.M. acknowledges financial support from the French Agence Nationale de la Recherche (ANR) through the ATTOMEMUCHO project (ANR-16-CE30-0001). S.Te. and S.B. are grateful for support through the Deutsche Forschungsgemeinschaft, project B03/SFB755 and project C02/SFB1073. P.J., S.M. and J.L. acknowledge support from the Swedish Research Council and the Swedish Foundation for Strategic Research. A.S.M. and P.K.O. acknowledge German-Russian Interdisciplinary Science Center (G-RISC, C-2015a-6, C-2015b-6 and C-2016b-7) funded by the German Federal Foreign Office via the German Academic Exchange
Service (DAAD) and Saint-Petersburg State University for financial support. A. Ro. is grateful for support through the Deutsche Forschungsgemeinschaft project RO 4577/1-1 and VR 76/1-1.
\end{acknowledgments}
\bibliography{TRPESCH3Iv1}{}

\end{document}